\documentclass[12pt,superscriptaddress]{revtex4}
\usepackage{graphicx}

\begin{document}

\title {Microsolvation of cationic dimers in $^4$He droplets: geometries of A$_2^+$(He)$_N$ (A=Li,Na,K) from
optimized energies.
}

\author{F. Marinetti}
\affiliation{Department of Chemistry, The University of Rome La Sapienza, Rome, Italy.}
\author{Ll. Uranga-Pi\~{n}a}
\affiliation{Department of Theoretical Physics, Physics Faculty,
University of Havana, Cuba.}
\author{E. Coccia}
\affiliation{Department of Chemistry, The University of Rome La Sapienza, Rome, Italy.}
\author{D. L\'opez-Dur\'an}
\affiliation{Instituto de Matem\'aticas y F\'isica Fundamental (CSIC), Madrid, Spain.}
\author{E. Bodo}
\affiliation{Department of Chemistry, The University of Rome La Sapienza, Rome, Italy.}
\author{F. A. Gianturco}
\altaffiliation[Corresponding author:]{$\quad$fa.gianturco@caspur.it. Fax: +39-06-4991.3305.}
\affiliation{Department of Chemistry, The University of Rome La Sapienza, Rome, Italy.}

\begin{abstract}
Ab initio computed interaction forces are employed in order to describe the microsolvation of
the A$_2^+(^2\Sigma)$ (A=Li,Na,K) molecular ion in $^4$He clusters of small variable size.
The minimum energy structures are obtained by performing energy minimization
based on a genetic algorithm
approach. The symmetry features of the collocation of solvent adatoms
around the dimeric cation are analyzed in detail, showing that the
selective growth of small clusters around the two sides of the ion during the solvation
process is a feature common to all three dopants.
\end{abstract}

\maketitle

\section{Introduction}
The characterization of the structures and energetics of atomic and
molecular aggregates is a current challenge of both theoretical and
experimental research directed to attaining accurate descriptions of
their nanoscopic properties. Weakly bound clusters of variable size,
especially those involving helium and other noble gas atoms, represent
an ideal testing ground for a large number of theoretical and
computational approaches \cite{01,02,03,04,05}: this is consequence of the fact that the
accurate knowledge of the relevant intermolecular forces between the
solvent atoms and the dopants present in the cluster is an important
prerequisite for the structural calculations and therefore fairly simple
components provide ideal model systems for the analysis of the
influence of intermolecular interactions on the cluster properties \cite{1,2,3}.

In the case of weak dispersion interactions, the ground state configuration is responsible for
most of the microsolvation effects on chemical species and determines the
outcome of possible chemical processes, which can in turn be used
to help the synthesis of new molecules and isomers \cite{4}.
For noble gas atoms this has become an active area of research
since, among them only the chemical properties of krypton are fairly
understood. Extensive theoretical analysis of the bonding in
compounds involving noble gases has been successful in that even
some new molecules have been predicted through theory \cite{5}.

From the experimental point of view, atomic or molecular impurities
embedded in noble gas clusters act as cromophoric units, allowing
their spectroscopic investigation. Information about the
structure of the cluster follows from the comparison between the
results of measurements in the cluster and the isolated cromophore
spectroscopy. Such an approach  makes possible the evaluation of the
effects of geometric modifications and of the solvent atoms
vibrations on the local excitation spectra of the impurity. Doped
helium clusters present some additional interesting features like
the rapid heat transport generated inside the complex to the
surface: due to such cooling, the He clusters are ideal candidates
for low temperature chemistry studies since the density of states
accessible to guest molecules is significantly reduced, minimizing
the complexity of the system especially from the theoretical point of view.
Helium droplets provide weakly-interacting, low temperature
environments suitable for studying the evolution of molecular
properties as the size of the cluster increases in a controlled
fashion. It allows to establish a continuous bridge between the
structural properties, energetics and chemical reaction dynamics in
isolated molecules and those of condensed phase. Furthermore, the
possibility to understand, at the microscopic level, the solvation
process  of an ionic  impurity has also motivated  a large  number
of theoretical and experimental research on ionic clusters involving
noble gas atoms. Impurities like positive ions form a region of
increased density due to electrostriction, leading to a larger local
density of the He adatoms and to the formation of short-range order.
This is known as the snowball model \cite{14} and has been widely
used in the interpretation of experimental data since this
phenomenological picture explains the low mobility of a positive ions
observed in experiment as compared with that of neutral species
\cite{15}. Therefore, positive ions are surrounded by many He atoms
that are strongly compressed as a result of electrostriction, so
that the resulting core is thought to lead to  solid-like structures
\cite{16,17}, i.e. to a \emph{snowball}. More specifically, the
analogy with a \emph{snowball} indicates the existence of some
compact microstructure surrounded by a more liquid-like shell of
less strongly localized $^4$He adatoms until the next solid-like
configuration is formed.

For various doped helium clusters, numerically converged calculations of
their structures and energies has been performed  by Variational
Monte Carlo (VMC) \cite{6} and Diffusion Monte Carlo (DMC) \cite{7,8}
techniques for the smaller aggregates, while for larger systems (N$\ge$100) the
full quantum mechanical approaches are still largely out of reach.
Even approximate methods are difficult to implement due to their
 computational cost, so that simplifying assumptions have to be
made in order to extend the  analysis to the larger aggregates.

Global optimization of  the bound structure energetics can provide a
suitable alternative to overcome such difficulties, a reason why
this technique may play a role in the determination of the
geometries of a wide variety of systems, such as proteins, crystals
and clusters \cite{9}. However, it also becomes an extremely costly
task as the number of atoms increases due to the exponential growth
of the local minima of comparable value: a great deal of effort has
therefore gone into developing efficient methods to find the lowest
energy structures of molecular clusters \cite{10}.

The interest on the interaction  between alkali metal atoms (Li, Na,
K) and rare gas atoms has also been renewed due to recent
experiments measuring the index of refraction of an atomic matter
wave passing through a dilute medium of rare gas atoms \cite{11}.
Clusters composed  by lithium and helium atoms therefore constitute
systems amenable to testing the effect of intermolecular
interactions on their structure since modern ab-initio and
model potential calculations are expected to provide
accurate descriptions of these interactions forces owing to the
small number of electrons involved in their description. Experimentally, the alkali
atom can be added  to the initial helium droplet by
taking advantage of  the ease with which these aggregates can pick
up one or more of such impurities \cite{12}, while the advent of the
more recent transmission grating observations have opened the
possibility to investigate the relative stability of weakly bound
clusters up to a few tens of helium atoms \cite{13}.

When ionization occurs, there is still a considerable lack of realistic microscopic
information about the solid order thought to be originated by the presence of the
ion \cite{18} and  a limited description of its dependence on
the chemical features in the case of molecular ionic dopants. In the
case of A$_2^+$(He)$_n$ (A=Li, Na and K) the occurrence of much deeper
attractive wells with respect to neutrals,
together with the effect of the potential anisotropy,  is expected to lead to
substantially different structures when compared with the neutral
counterparts.

The aim of this work is therefore that of employing a genetic
algorithm approach for the study of the ground state  of
A$_2^+$(He)$_n$ clusters, an approach which relies on the energy optimization
on the global potential energy surface of the cluster.

There are obviously questions related to such an approach in
relation to the quantum nature of the nanoaggregates where Van der Waals (VdW) forces
are expected to dominate the
interaction landscape. Thus, we have analyzed in the past the reliability
of a classical picture for describing ionic dopants in a quantum
solvent like bosonic He \cite{19,20,21} and found that the presence
of the stronger ionic interactions drives the overall stability of
the smaller aggregates and causes two chief effects in such systems,
as already mentioned before (i.e. electrostriction and snowballs), which
contribute to  confirm the classical results.

Furthermore, the selection of a simplified description of the
overall interaction which we model via the sum-of-potentials (SOP) scheme
\begin{equation}
V(\mathbf{R},\mathbf{r})=\sum_{i=1}^N
V_\mathrm{He-M}(\mathbf{r}_i)+\sum_{ij}V_\mathrm{He-He}(\mathbf{R}_{ij})~,
\end{equation}
where the vectors \textbf{r}$_i$ label the interactions
between each $^4$He solvent atom and the dopant molecule M, while
the \textbf{R}$_{ij}$'s describe the distances between the helium
adatoms within each cluster of size N, allows us to immediately relate the features
of the component two-body (2B) potentials   with the behavior of each cluster as a
function of its size.

In the present systems, as done before by us \cite{22}, we have included an
additional correction to account for the effects due to many-body (MB) forces. The latter,
however, are found not to change the overall pattern of
structural evolution produced by the SOP scheme while slightly changing the actual energy values.

The paper is organized as follows: in  Section II the method and the
potential energy surface employed in this work are briefly
reviewed while in Section III we present the main results obtained
on the energetics and the spatial distributions. Finally, in
Section IV our conclusions are summarized.

\section{The theoretical machinery: an outline}
To obtain an analytical expression for the total cluster energy, we
start by adding only the contributions coming from the impurity-helium \cite{22}
and helium-helium potentials \cite{23}, thereby initially neglecting the
many-body effects present in the clusters and following the SOP
scheme of eq. (1).  A large number of previous studies have been
devoted to the evaluation of the leading three-body contributions to
interaction potentials between several complexes and noble gas
atoms (see for instance our work \cite{24} and references therein), since besides
their contributions to the binding energy, three-body terms
 may in principle influence the geometry of the cluster.
In spite of their limited importance found by
our previous calculations on this type of systems \cite{19,20,21},
we have incorporated them through the procedure described below.

The chief reason why such many-body effects  remain fairly small in comparison with the total
energy \cite{20,21,22,23,24,25} is  because of the small polarizability
of its closed-shell configuration where no significant changes
are expected to occur in the electronic structure of helium atoms
due to the presence of the ionic impurity. Thus, the only MB correction that we have considered
in our calculations is the coupling between the dipole moments induced by the
cation on the electron density of a pair of helium atoms.  The analytical
formula for this term is very simple once we consider a point charge
located at the center of mass of the molecular ion, and is given by
\begin{equation}
V_{3B}(r_{ij},r_i,r_j) = - \frac{\mu
_i\mu_j}{r_{ij}^3}\left[2\cos\vartheta_i\cos\vartheta_j-\sin\vartheta_i
sin\vartheta _j\right]~,
\end{equation}
where the angles are formed between each dipole distance
from the point like charge ($r_i,r_j$) and the line joining them
($r_{ij}$). Each dipole moment $\mu _i$ can then be evaluated via
the well-known formula: $\mu _i=\alpha /r_i^2$, $\alpha$ being the
He polarizability.

The interactions between each
solvent atom and the molecular dopant has been the subject of
previous studies in our group, where the He-A$_2^+$ anisotropic
potential energy surfaces have been evaluated with the ionic
core molecular bond kept at its equilibrium value \cite{22}.

\begin{figure}
\includegraphics[width=0.8\textwidth]{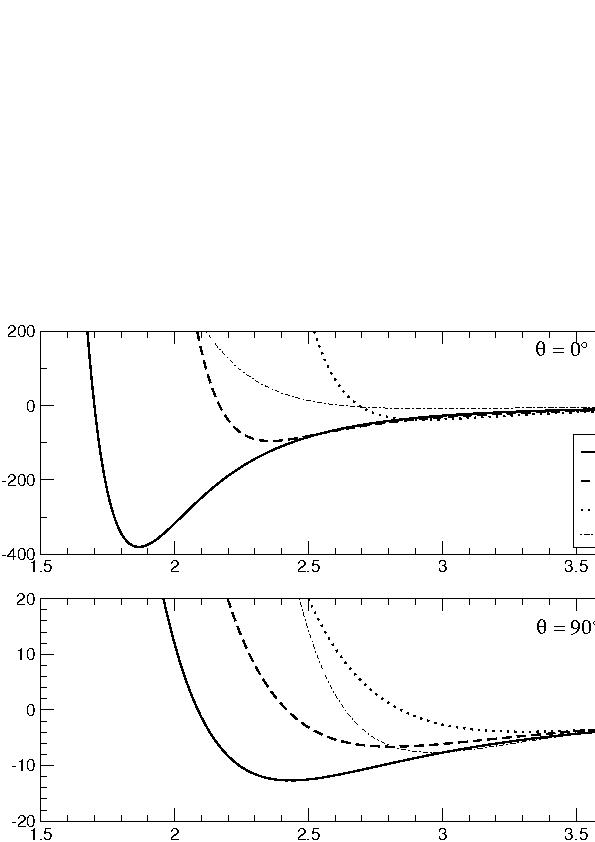}
\caption{A$_2^+$-He interaction as obtained from the ab initio calculations
of \cite{22}: units in \AA and cm$^{-1}$. The dash-dotted line is the He-He potential
from ref \cite{23}}\label{pot}
\end{figure}

The relevant features of the intermolecular potentials associated
with a single $^4$He atom can be appreciated by looking at the data of  figure \ref{pot}:
the strongly anisotropic character of the angular dependence of the
PES is clearly seen for all the three dimer ions. The absolute minima for the three PESs are collinear
while they also present an angular saddle in the C$_{2v}$ geometry, where each
potential is much less attractive. The maximum well depths are -380, -95 and -37 cm$^{-1}$ for
Li$_2^+$, Na$_2^+$ and K$_2^+$ respectively.

In the same figure we report also the He-He potential: its relative strength with respect to the He-ion one
has an important influence on the growth of the cluster: as we will show below, since Li$_2^+$-He is the strongest
interaction, the collocation of the He adatoms will occur on both sides of the molecule almost
independently. In the case of K$_2^+$, instead, the less attractive ion-He interaction is not able
to balance the weak He-He potential and the adatoms prefer to arrange themselves
asymmetrically on the two linear and equivalent well regions.

The details of the  minimization method has been
given previously \cite{26} and therefore we will not repeat them
here: in what follows we will only outline our present approach. The
classical optimization is based on the well-known genetic algorithm
\cite{10,27}, a procedure which can provide a reasonably fast route
to finding the minima for a general function. The method employed in
the present study follows closely the scheme described in ref.
\cite{10}.

In spite of the marked localization effects of the He adatoms of the
inner shells which is caused by the ionic dopant \cite{28}, the
quantum nature of the present solvent should also be considered when
analyzing the final spatial structures. We take into
account  such effects by  substituting the $\delta$ functions of the
classical picture of atomic network locations  by finite spread distributions along the radial
variables. A Gaussian function is thus employed to simulate the zero
point vibrations of each cluster particle. As a consequence of it,  the corresponding radial
densities can be approximated through the following expression:
\begin{equation}
 \rho(r)=N\sum_{i=1}^{N}\frac{1}{\sqrt{(2\pi\sigma_i^2)}}
  \exp\left(-\frac{(r-\bar{r}_i)^2}{2\sigma_i^2}\right)\label{dist}
\end{equation}
where $\sigma_i$ is chosen as the product of the standard deviation of the
quantum ground vibrational state of the Li$_2^+$-He system, already
computed earlier by us \cite{22}, and the uncertainty on the radial values
obtained from the $\eta _j^i$ parameters of the genetic algorithm. The r$_i$ is the i-th atom distance
from the center-of-mass of the impurity after optimization.
This analysis allows us to obtain spread distributions for
the clustering  of the \emph{classical} He atoms around the dopant
ion, hence by somehow including quantum delocalization effects for the adatoms near the ionic dopant.

We shall see below how such corrections do includes delocalization but
follows qualitatively the finding from the classical description
provided by our optimization scheme, the latter being however much more helpful
in describing  spatial collocations of the solvent atoms
within the shells closer to the ionic dopants.

\section{Energetics and structures}
The presence of long-range polarization forces causes the effects
of the impurity-helium interactions to
extend over a considerable amount of solvent atoms within the cluster structure.
We have therefore analyzed  a broad range of cluster sizes for a nonvibrating
molecular partner in order to unravel such effects in the title
systems and to relate them to their specific interaction potentials.

\begin{figure}
\includegraphics[width=0.8\textwidth]{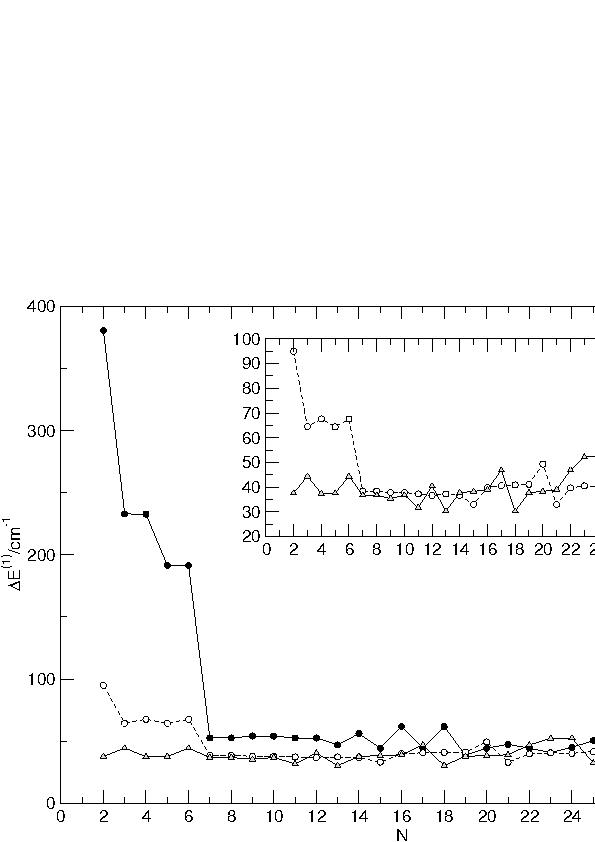}
\caption{Computed evaporation energies of the helium clusters with
the three cationic dopants.}\label{evap}
\end{figure}

To begin with,
we report in figure \ref{evap} the single-atom
evaporative energies which are defined as:
\begin{equation}
\Delta E_N^{(1)} =E_{N-1}-E_N~.
\end{equation}
For all the three type of clusters we see some interesting features. The pronounced
steps appearing in the evaporation energy in the small cluster range are a
signatures of the completion of some solvation sub-shells, while the peaks are imprints
of the appearance of particularly stable structures (magic numbers). For larger clusters
those energies present a slightly oscillatory pattern as additional helium atoms are added during the
system's growth: the mean value is approximately 50 cm$^{-1}$ for all the three systems.
The K$_2^+$(He)$_N$ case however, does not show any clear indication of completion
of a solvation shell, but only of some  ``magic'' numbers appearing at $N$=6, 12, 17, and 24.
For the Na$^+_2$ case we do see the filling of the first shell with 6 adatoms
and a magic number at $N$=20. In the case of the Li$_2^+$ the structuring of the first shell is evident at $N$=6 and three small  peaks at $N$=14, 16, 18 are signatures of additional stable species.
It is worth pointing out here that for both sodium and lithium ionic dimers the first two He atoms are strongly bound
to either sides of the impurity as independent partners.

\begin{figure}
\includegraphics[width=0.7\textwidth]{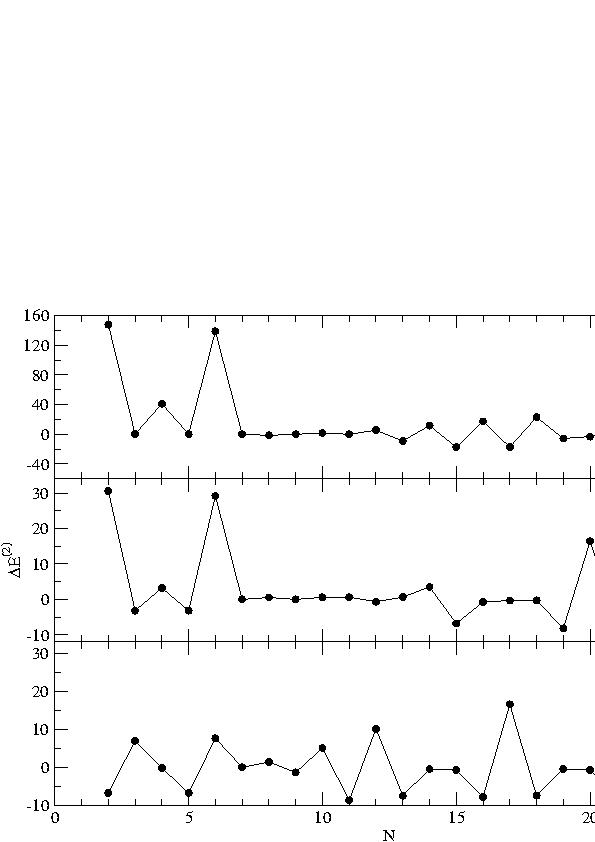}
\caption{Computed 2$^{nd}$ order differences for the three
alkali-metal ionic dimers as functions of the no. of He atoms in the
clusters.} \label{evap2}
\end{figure}

These effects are even more clearly appreciable in the behavior of the
second energy differences reported by figure \ref{evap2}. This indicator measures the
stability of each cluster with respect to the nearest ones and is
 useful for the detection of shell structures or of ``magic''
numbers when observing cluster growth. It  is defined as
\begin{equation}
\Delta E_N^{(2)}=E_{N+1}+E_{N-1}-2E_N~.
\end{equation}
The data reported in figure \ref{evap2} confirm substantially our previous findings in terms
of magic numbers.

\begin{figure}
\includegraphics[width=0.8\textwidth]{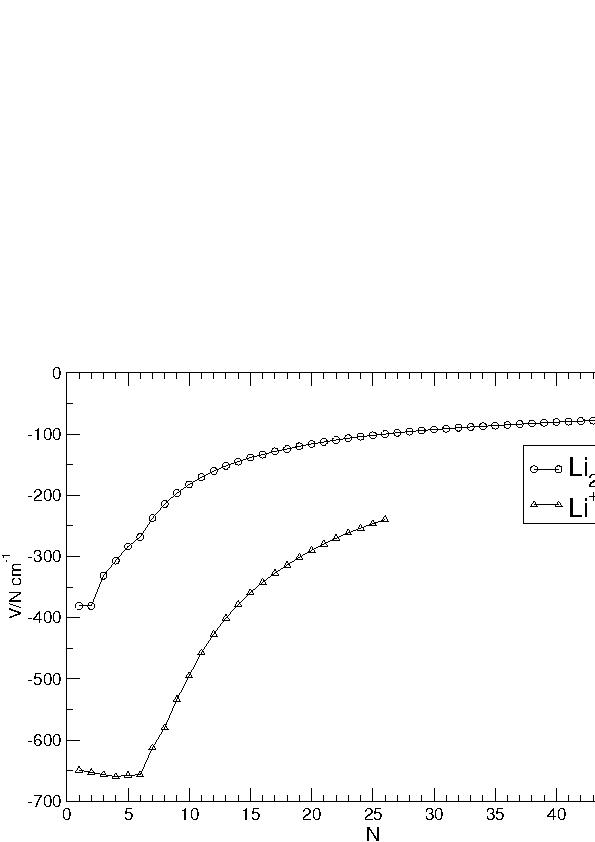}
\caption{Computed average energy per added solvent atom. The
calculations for
 Li$^+$ are reported in ref. \cite{29}).}\label{evap3}
\end{figure}

From the data related to the intensive quantity E$_{N}/N$
 and reported in figure \ref{evap3} only for the Li-doped  species, one can see
that the average energies for Li$^+_2$ (open circles) tend to reach an asymptotic value of about $75$ cm$^{-1}$ as the
cluster size increases. Beyond this value, in fact, the inclusion of any new adatom
to the cluster produces almost the same effect on the energetics.
On the other hand, it is clear from the initial values of
this quantity that only the first two $^4$He adatoms can be regarded
as truly independent partners while the additional ones, although still
strongly bound to the cation, show the helium-helium interactions
important role in reducing the incremental contributions of sequential adatoms in the small and
intermediate cluster sizes.
It is interesting to compare such data with similar calculations for $^4$He
clusters containing the cationic monomer Li$^+$ as a
dopant \cite{29}; they yield the E$_{N}/N$ values reported in the
same figure \ref{evap3} by open triangles. The number of solvent atoms which
attach to the Li$^+$ ion in a nearly independent manner is now
6 while, on the other hand, we clearly see the same overall trend as that shown
by the dimer in terms of changes of the E$_{N}/N$ binding energy as
$N$ increases: the latter quantity is progressing towards an asymptotic value as in
the dimer case. In the larger clusters of both species the adatoms will
view the dopant essentially as a localized, point-like positive charge.

\begin{figure}
\includegraphics[angle=180,width=0.8\textwidth]{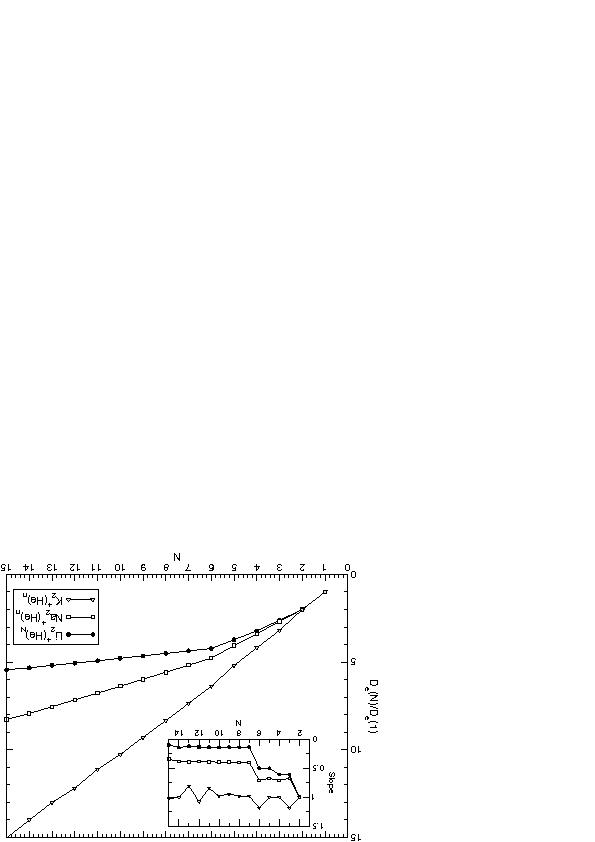}
\caption{$D_e(N)/D_e(1)$ ratios for the three dopant
species.}\label{evap4}
\end{figure}

Another way of looking at the solvation process described by our classical calculation is to consider the ``normalized'' total and evaporation energy: we report in figure \ref{evap4} the classical minima energies divided by the $D_e$ of the single He moiety. In the inset we also show the slope of such quantities with respect to $N$. The growth of the cluster for Li$_2^+$ and Na$^+_2$ is clearly  marked by the existence of a 6-atom shell and by an almost constant rate of energy acquisition during the accretion process.
Above $N=6$, for Li$_2^+$ and Na$_2^+$ the rate of increase in binding energy is of about 0.2$D_e$ and 0.4$D_e$ respectively for each He atom. The case of K$_2^+$, on the other hand, does not show any special structuring at $N=6$ and the  rate of increase in energy is 1.0$D_e$, a value which shows a behavior analogous to that of a pure He cluster, albeit on a different scale, where the same energy is gained for each adatom attached to the others. We conclude that, despite  being a cationic molecular ion, the solvation process around K$_2^+$ behaves as if it were one of the  equivalent He atoms so that  from an energy point of view, its role is not different from any other adatom in the cluster.

\subsection{A selection of clusters' geometries}
\begin{figure}
\includegraphics[angle=-90,width=0.8\textwidth]{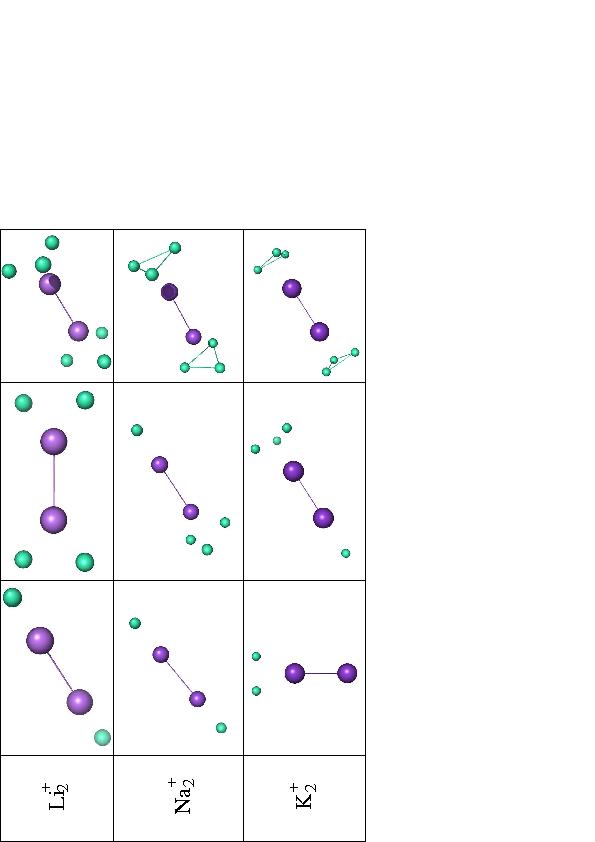}
\caption{Optimal spatial structures of the clusters of Li$_2^+$,
Na$_2^+$ and K$_2^+$ for N=2, 4, 6. The darker, larger spheres
represent the cationic dopant.}\label{struct1}
\end{figure}

The geometries of the optimized clusters for Li$_2^+$, Na$_2^+$ and K$_2^+$ for $N$=2, 4 and 6 are shown in figure \ref{struct1}. The pictorial story told by that figure clearly indicates that for the Li$_2^+$  case the solvation proceeds  symmetrically at the two ends of the  molecule. On the other hand, in the Na$_2^+$  the resulting structures are less symmetric because of the competing  energy gain coming from forming  He-He networks on one side of the molecular ion.   For the case of K$_2^+$ the He atoms prefer to locate themselves on one side only:
this is obviously due to the weakening of the dimer-He interaction which then
becomes competitive  with the He-He potential giving rise to the preferential formation  of the mentioned He-He networks.

\begin{figure}
\includegraphics[angle=-90,width=0.8\textwidth]{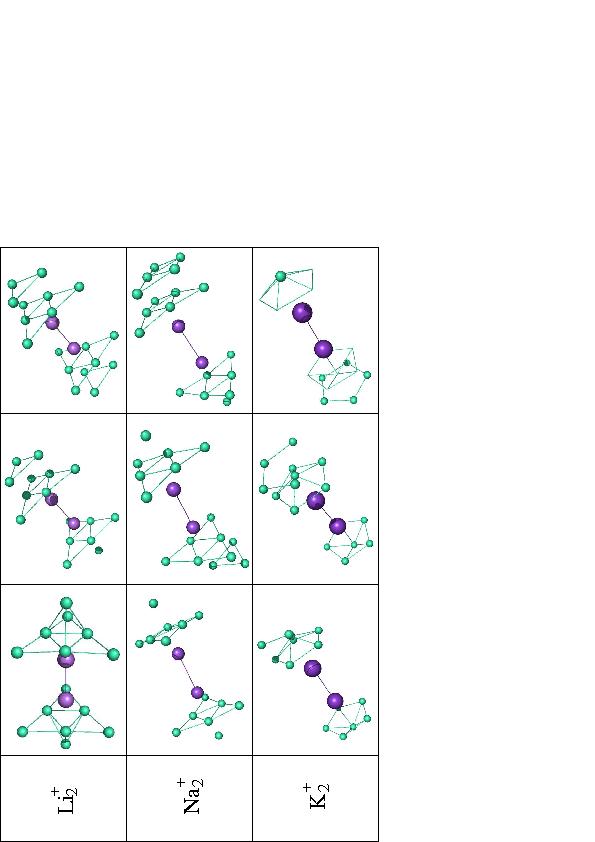}
\caption{Structures of the optimized geometries for the
A$_2^+$(He)$_N$ clusters with $N$=14, 16, 18.}\label{struct2}
\end{figure}

Figure \ref{struct2} further shows a selection of optimized geometries for the larger clusters:
we see that triangular, tetrahedral and planar configurations are
formed almost independently around each cationic center. The dominance of
linear wells in the Li$_2^+$-He interactions pushes the sequential
adatoms to be collocated in them as close as possible to each other
so that the two nearly independent microsolvation \emph{cages}  grow around each of the Li
atoms in the Li$_2^+$ dimeric cation. This independence effect does not occur for clusters which contain Na$_2^+$ and K$_2^+$ as dopants since, due to their weaker interactions with each solvent atom, experience a marked competition with the clustering process of He atoms.

\subsection{The quantum effects}
In order to provide an estimate of the quantum effects  due to delocalization and ZPE, we have performed accurate Variational and Diffusion Montecarlo calculations on the smaller moieties containing all  the three species.
The method has been extensively given before by us \cite{28} and therefore will not be described again in the present work.
We report in Table I the quantum energies ($D_0$), the classical minima ($D_e$) and the ZPE \%.

\begin{table}
 \begin{tabular}{ccccc}
$N$ & $D_0(N)$ & $D_e(N)$ & ~~\%ZPE~~ &  $D_0(N)/D_0(1)$ \\
\hline
\multicolumn{5}{c}{Li$_2^+$} \\
\hline
1 & -254.081 $\pm$ 0.035 & -380.945 & 33.0 & 1.0 \\
2 & -515.750 $\pm$ 0.130 & -761.781 & 32.3 & 2.0 \\
3 & -639.104 $\pm$ 0.800 & -994.777& 35.7 & 2.5 \\
4 & -895.998 $\pm$ 0.100 & -1227.588 &27.0  & 3.5 \\
\hline
\multicolumn{5}{c}{Na$_2^+$} \\
\hline
1 & -49.462 $\pm$ 0.340 & -95.026 & 47.9 & 1.0\\
2 & -98.334 $\pm$ 0.120 & -190.033 & 48.2 & 2.0\\
3 & -127.810 $\pm$0.056 &-254.476 & 49.8 & 2.6\\
4 & -151.076 $\pm$0.120 & -322.083& 53.1& 3.0\\
\hline
\multicolumn{5}{c}{K$_2^+$} \\
\hline
1 & -17.470 $\pm$ 0.056 & -37.41 & 53.3 & 1.0\\
2 & -35.020 $\pm$0.073 & -75.01& 53.3& 2.0\\
3 & -47.930$\pm$0.149&-119.40 & 59.9& 2.7\\
4 &-60.770 $\pm$0.188&-156.80 &61.2 & 3.5\\
\hline
\hline
 \end{tabular}
\caption{Classical ($D_e$) and quantum ($D_0$) binding  energies for the smaller clusters. All energies in wavenumbers.}
\end{table}

As can be seen from the data reported in Table I, and especially from those in its last column, the quantum results substantially confirm the picture of an  independent growth of the cluster around the two sides of the molecule (see for comparison figure \ref{evap4}). The ratios  $D_0(N)/D_0(1)$ for Li$_2^+$ and Na$^+_2$ are in fact consistent with the classical result of a symmetric building process which accommodates two atoms by the two molecular sides and then continues building the cluster by alternatively placing the second and the third atom. The quantum results for K$_2^+$  are however different from our classical findings since the quantum solvation seems to proceed symmetrically when correctly including quantum effects. This  is related to the large ZPE values existing in K$_2^+$(He)$_n$ which therefore makes more stable the symmetric structures such as $N=4$ with an even number of adatoms.

\begin{figure}
\includegraphics[width=0.8\textwidth]{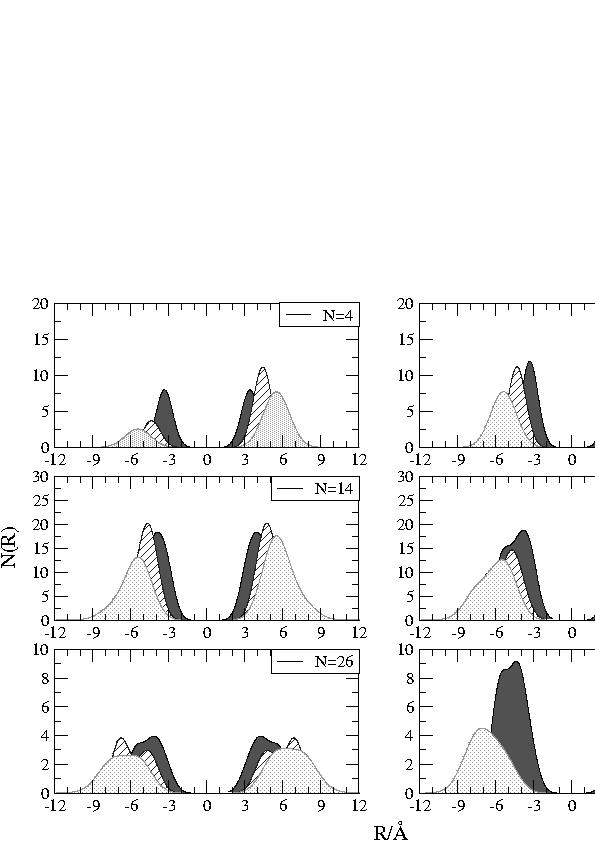}
\caption{Computed radial distributions of solvent atoms from the
dopant's center-of-mass for different cluster sizes. Li$_2^+$,
Na$_2^+$ and K$_2^+$ distributions are represented  by the
differently shaded areas: dark for Li$_2^+$, lines for Na$_2^+$ and
gray for K$_2^+$.}\label{distr}
\end{figure}

In order to also account for the effects of delocalization on the classically optimized structures, we report in figure \ref{distr} the radial distributions of the helium atoms on both sides of the molecules. We have generated such distributions starting with the classical results while the  final, ``quantum''  distributions have been obtained from  eq. (\ref{dist}) using the standard deviation inferred from a ground state DVR wavefunction for the impurity-He system as discussed in our earlier work \cite{22}.
Since the solvation cage is often asymmetric with respect to the two end of the cationic dimer, we have counted the number of particles of each side of the molecules and we have normalized the left-hand and right-hand side of the density accordingly.

As we can see in figure \ref{distr}, the Li$_2^+$(He)$_N$ clusters are almost symmetric (for an even number of He atoms) up to $N$=18, the Na$_2^+$ doped ones have both symmetric and asymmetric geometries. Finally the K$_2^+$ doped clusters have clearly asymmetric spatial arrangements.

\section{Conclusions}
The data reported in this work have focused on  the energetics and the
spatial collocations of solvent atoms in  $^4$He droplets doped by three different
cationic dimers:  Li$_2^+$, Na$_2^+$ and K$_2^+$.

The interactions within the clusters have been obtained by a sum of the two-body potentials acting between its
components and derived  from ab initio calculations \cite{22}. A realistic approximation to the three body term
has also been included in the energy expression to evaluate its effects which have turned out to be rather small.

The optimization procedure we have employed here is based on a genetic algorithm
scheme already discussed by us before \cite{25,26} and the
optimized structures have been analyzed in terms of their geometric features
in order to explain  the increased asymmetry of the solvation shells around the cationic dimer
when going from Li$_2^+$ to K$_2^+$ as dopants.

The following characteristics were found to occur along the series of
the examined clusters: in the clusters doped by Li$_2^+$ and Na$_2^+$, the interaction
between each He atom and the ion is much stronger than the He-He potential and therefore  causes
the formation of quasi-linear accumulation cages of solvent atoms with respect to the molecular  dopants.  The initial solvent particles thus start to place themselves at the two ends of the molecule and far away from each other, thereby creating an initial subshell of 6 atoms.
Adding more He atoms makes the two parts of this subshell grow in an independent fashion
by collocating solvent atoms alternatively on either of the two sides of the dimer.

The present study, therefore confirms the expected dominance
of ionic forces in driving cluster shapes when the dopant is a
cationic molecule and clearly shows the crucial role played by changing the competitive strength of the ion-He interaction with respect to the He-He network of interactions. Thus, for the  K$_2^+$ dimer we found  that the construction of the solvent cage  is asymmetric and dominated by the  He-He network formation. For Li$_2^+$  and
Na$_2^+$, on the other hand,  we have  shown that a rigid \emph{snowball} structure is
formed as a regular configuration within the first
shell that encloses the solvated cation (with $N$=6), followed by
more delocalized collocations of the solvation atoms for clusters
beyond that size.  Such collocations are largely creating symmetric solvent cages at the two ends of each ionic dimer.

In spite of using a classical representation of the solvent atoms, we have shown that for ionic molecular dopants (as it has occurred with atomic ions \cite{19}) the final description of the microsolvation process in the smaller He clusters turns out to be keeping with the quantum results and indicates the present approach as providing useful insights on that process.

\section{acknowledgments}
The financial support of the Research Committee of the University of
Rome "La Sapienza" is gratefully acknowledged. L. U.
thanks the Department of Chemistry of the University of Rome ``La Sapienza''
for the award of a research fellowship during which this work was completed, while D. L.-D.
thanks the European Community for financial support.

The computational support of the CASPUR Supercomputing Consortium is
also acknowledged. Finally we also thank the support of the INTAS project.

\end{document}